\documentclass{article}
\usepackage{LaThuileFPSpro}
\usepackage{graphicx} 
\usepackage{dcolumn}  
\usepackage{amsmath, amsthm, amssymb} 

\graphicspath{{ps/}}

\newcommand{\ra}{\rightarrow}

\newcommand{\ks}{K_S}
\newcommand{\lamst}{\Lambda(1520)}
\newcommand{\tht}{\Theta^+}

\newcommand{\thc}{\Theta_c^0}

\newcommand{\cm}{\mathrm{cm}}

\newcommand{\kp}{K^+}

\begin{document}
\title{EXPERIMENTAL REVIEW ON PENTAQUARKS}

\author{Michael Danilov \\ 
{\em Institute for Theoretical and Experimental Physics}\\
{\em B.Cheremushkinskaya 25}\\
{\em 117218 Moscow}\\
{\em Russia}
}

\maketitle

\vspace*{0.2cm}
\begin{center}
Talk at the Les Rencontres de Physique de la Vallee d'Aoste\\
\it{(February 2005)}
\end{center}

\vspace*{1.2cm}

\baselineskip=11.6pt

\begin{abstract}
The experimental evidence for pentaquarks is reviewed and compared with 
the experiments that do not see any sign of pentaquarks.   
\end{abstract}
\newpage
\section{Observation of pentaquarks}
\begin{figure}[tbh]
\centering
\begin{picture}(550,140)
\put(10,-15){\includegraphics[width=11cm]{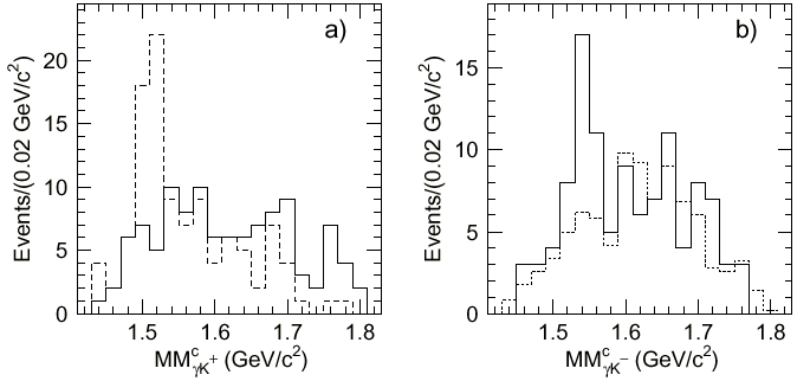}}
\end{picture}
\caption{\it Missing mass spectra for the $\gamma K^+$ (left) and $\gamma K^-$ (right) for the reaction
$\gamma C\ra K^+K^-X$~\cite{leps}. The dashed (solid) histogram shows events
 with (without) additional detected proton.
The $\lamst$ signal is seen on the left and evidence for $\tht$ is seen on the right.} 
\label{LEPS}
\end{figure}

\begin{figure}[tbh]
\centering
\begin{picture}(550,120)
\put(60,-10){\includegraphics[width=7.5cm]{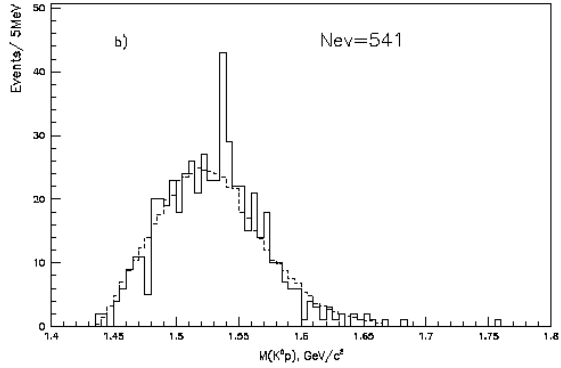}}
\end{picture}
\caption{\it Invariant mass of $pK^0$ in the reaction $K^+Xe\ra p\ks X$~\cite{diana}.
         The dashed histogram is the expected background.} 
\label{DIANA}
\end{figure}

 Until recently, all existing baryons could be interpreted as bound
 states of three quarks. 
Observations of a pentaquark state $\tht$ in $n\kp$~\cite{leps} and
 $pK^0$~\cite{diana} modes
 created a lot of excitement.
 The corresponding invariant mass distributions obtained by the
 LEPS and DIANA collaborations are shown 
 in Fig.~\ref{LEPS}~,~\ref{DIANA}.

 The minimal quark content of the $\tht$ is $uud d\overline s$. 
 Thus for the first time unambiguous evidence
 was obtained for hadrons with an additional quark-antiquark
 pair.

\begin{figure}[tbh]
\centering
\begin{picture}(550,150)
\put(40,-10){\includegraphics[width=8cm]{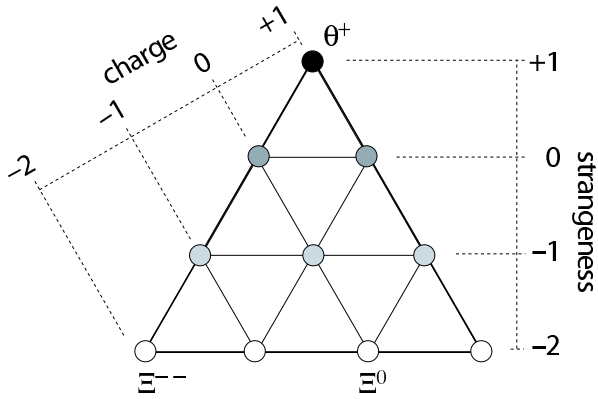}}
\end{picture}
\caption{\it The predicted anti-decuplet~\cite{dpp} of pentaquark baryons.
Experimental evidence for three indicated particles has been presented.} 
\label{antidecuplet}
\end{figure}

 Analysis of the DIANA data demonstrates that the width of the $\tht$  is
 very small $\Gamma=0.9\pm 0.3$ MeV~\cite{cahn}.
 A similar small width was obtained from the
 analysis of the $\kp d$ cross
 section~\cite{nussinov}~$^-$~\cite{gibbs}. 
Such a narrow width is
 extremely unusual for hadronic decays and requires reassessment of
 our understanding of quark dynamics. Properties of the $\tht$ were in the excellent
 agreement with the theoretical predictions~\cite{dpp} based on the chiral quark soliton
 model. This paper motivated both
 experimental searches although later on the accuracy of these
 predictions was questioned~\cite{ellis}.
 In the quark soliton model the $\tht$ belongs to an antidecuplet of baryons
 (see Fig.~\ref{antidecuplet}). 
 Octet, decuplet, 27-plet, and 35-plet of pentaquarks are also expected.

 Many experiments promptly confirmed
 the existence of the $\tht$~\cite{clas-d}~$^-$~\cite{lpi} in different processes:
 photoproduction, deep inelastic scattering, hadroproduction, and 
 neutrino interactions. Table~\ref{positive} shows properties of the observed peaks. 

There is some spread in the mass values obtained
 by different experiments. In particular masses in the $p\ks$ final state
 are lower than in the $n\kp$ one. The accuracy of the mass determination
 is not high in most of the experiments and therefore 
 the disagreement is not very serious 
 statistically. However the DIANA and ZEUS
 measurements are quite precise and contradict each other by more than
 4 sigma.  Several experiments observe finite width of the $\tht$  that is
 much larger than 1 MeV. However, the accuracy is again not high and
 within 3 sigma all width measurements are consistent with the
 instrumental resolution.  
\begin{center}
\begin{table}[ht]
\caption{\it Experiments with evidence for the $\tht$ baryon.}
\centering
\vskip 0.1 in
\begin{tabular}{|c|l|c|c|c|}
\hline
Reference& Group& Reaction 	& Mass 		& Width  \\
	&	&		& (MeV)		&(MeV)	  \\
\hline
\cite{leps} &
LEPS(1)	& $\gamma C \to K^+ K^- X$	& $1540\pm 10$	& $<25$	 \\
\cite{diana} &
DIANA	& $K^+ Xe \to K^0 p X$		& $1539\pm 2$	& $<9$	 \\
\cite{clas-d} &
CLAS(d)	& $\gamma d \to K^+ K^- p (n)$	& $1542\pm 5$	& $<21$	\\
\cite{saphir} &
SAPHIR	& $\gamma d \to K^+ {\overline {K^0}} (n)$	& $1540\pm 6$	& $<25$	 \\
\cite{itep} &
$\nu BC$	& $\nu A \to K^0_s p X$		& $1533\pm 5$	& $<20$	 \\
\cite{clas-p} &
CLAS	& $\gamma p\to\pi^+ K^+K^-(n)$	& $1555\pm 10$	& $<26$	 \\
\cite{hermes} &
HERMES	& $e^+ d \to K^0_S p X$		& $1526\pm 3$	& $13\pm 9$\\
\cite{zeus} &
ZEUS	& $e^+ p \to K^0_S p X$	& $1522\pm 3$	& $8\pm 4$\\
\cite{cosy} &
COSY-TOF& $p p \to K^0 p \Sigma^+$	& $1530\pm 5$	& $<18$	 \\
\cite{svd} &
SVD	& $p A \to K^0_S p X$		& $1526\pm 5$	& $<24$	  \\
\cite{leps2} &
LEPS(2) &          $\gamma d \to K^+ K^- X$ & $\sim 1530   $   & $  $   \\
\cite{itep2} &
$\nu BC2$	& $\nu A \to K^0_s p X$		& $1532\pm 2$	& $<12$	 \\
\cite{nomad}& 
NOMAD     & $\nu A \to K^0_s p X$       & $1529\pm 3$	& $<9$	 \\
\cite{jinr}& 
JINR     & $p(C_3H_8) \to K^0_s p X$       & $1545\pm 12 $	& $16\pm 4$	 \\
\cite{jinr2}& 
JINR(2)     & $CC \to K^0_s p X$       & $1532\pm 6$	& $<26$	 \\
\cite{lpi}& 
LPI     & $np \to npK^+K^-$       & $1541\pm 5$	& $<11$	 \\
\hline
\end{tabular}
\label{positive}
\end{table}
\end{center}
 
    The spread in mass and width may indicate that some experiments
 observe not a signal but a statistical fluctuation. 

    If the penaquark interpretation of observed peaks is correct one
 expects many other exotic (or cripto exotic) baryons
 belonging to the same antidecuplet or other multiplets. Indeed several
 experiments observe additional 
 peaks in the vicinity of the $\tht$ mass~\cite{itep2,jinr,lpi}.
 For example 3 peaks with
 the estimated statistical significance of 7.1, 5.0, and 4.5 sigma are seen in
 neutrino interactions~\cite{itep2}.
 
\begin{figure}[h!]
\begin{picture}(550,190)
\put(60,-10){\includegraphics[width=6.cm]{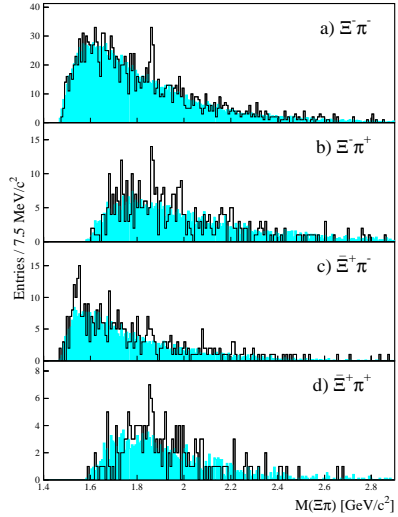}}
\end{picture}
\caption{\it Invariant mass spectra for $\Xi^-\pi^-$ (a),$\Xi^-\pi^+$ (b),$\overline{\Xi}^+\pi^-$ (c),
and $\overline{\Xi}^+\pi^+$ (d) in the NA49 experiment. The shaded histograms are the normalized
mixed-event backgrounds.} 
\label{NA49}
\end{figure}

    The NA49 collaboration claims an observation of a double strange
 pentaquark~\cite{na49}.
 Two observed narrow resonances $\Xi^{--}_{\overline {10}}$ and $\Xi^{0}_{\overline {10}}$  (see
 Fig.~\ref{NA49} )
fit naturally into the same antidecuplet as the $\tht$
 (see Fig.~\ref{antidecuplet}).

\begin{figure}[h!]
\begin{picture}(550,110)
\put(40,-30){\includegraphics[width=7.0cm]{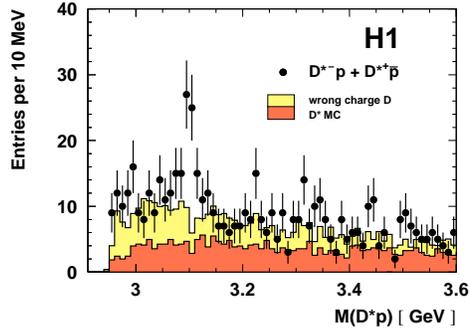}}
\end{picture}
\caption{\it Invariant mass distribution of $D^{*-}p$ and $D^{*+}\overline p$ combinations in the H1 experiment.
Two background components are shown as the shaded histograms.} 
\label{H1}
\end{figure}

An evidence for an
 anti-charmed pentaquark was obtained by the  H1 collaboration~\cite{h1} (see
 Fig.~\ref{H1}).



\section{Reliability of pentaquark observations}
  
    The evidence for pentaquarks was criticized by several authors
 (for a review see ~\cite{dsierba}). They considered kinematic reflections,
 ghost tracks and arbitrary selection criteria as possible
 explanations for the observed peaks. The first two worries were shown
 to be not important at least in some experiments (for a review see~\cite{hicks}).
 The last point
 is especially serious since statistical significance of the positive
 experiments is not high and thus they are vulnerable to a psychological
 bias. This problem is illustrated by the JINR analysis ~\cite{jinr}
 in which authors without any reason discard the momentum range where they
 do not see the signal
. The ZUES collaboration does
 not see the signal in data with $Q^2<20$ GeV$^2$. Their justification for
 discarding these data is also not too convincing. There are other
 examples of experiments with not well justified cuts. On the other
 hand there are experiments (for example DIANA) in which event
 selection criteria have high efficiency and reasonably justified. 

\begin{figure}[h!]
\centering
\begin{picture}(550,330)
\put(0,-10){\includegraphics[width=12cm]{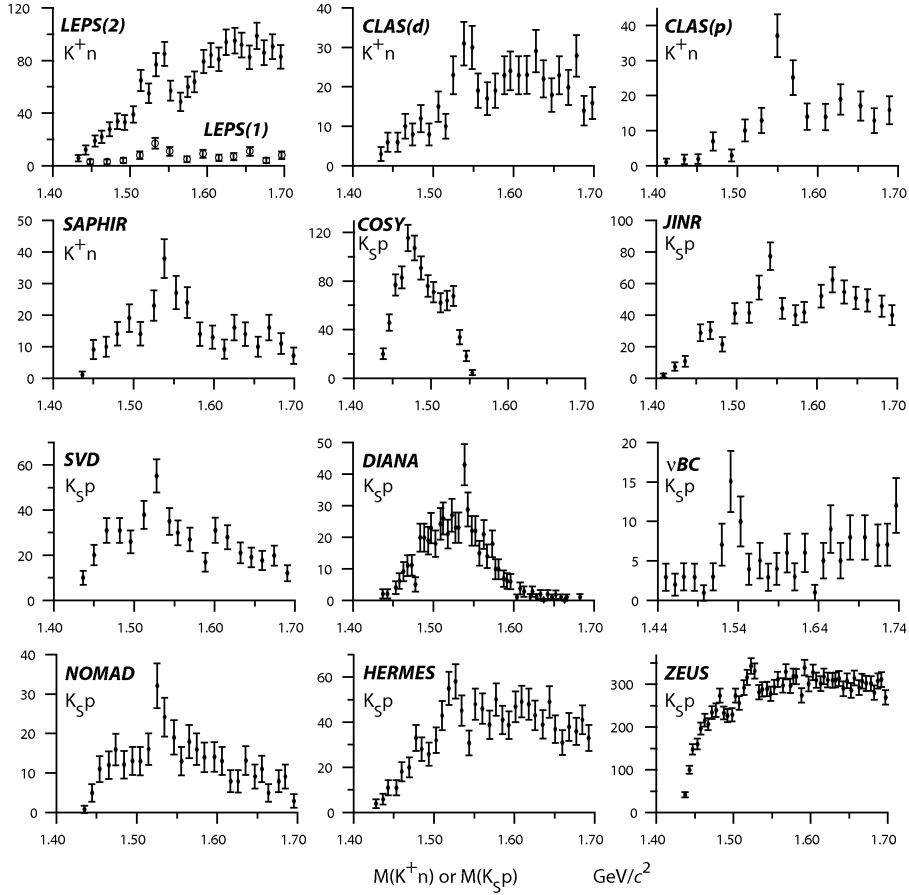}}
\end{picture}
\caption{\it Mass spectra of $nK^+$ 
 and $p\ks$ pairs in the experiments which provide evidence for the  $\tht$.} 
\label{all}
\end{figure}

The statistical significance of peaks is overestimated in all experiments
 since the shape of the background is not known. This looks obvious if
 one removes the fit curves and plot the data points with error bars
 (see Fig.~\ref{all} taken from \cite{dsierba}). 

Nevertheless the number of experiments is large and the combined
 significance is high if we disregard for a moment the spread in the
 peak position and width.  
    So one can not prove that all observed peaks are fakes or statistical
 fluctuations. Only high statistics experiments can confirm or disprove
 the claim for pentaquarks.

\section{Non-observation experiments}
    
    Experiments which do not observe pentaquarks are shown in Table~\ref{negative}.
 Many of them are high statistics experiments which observe by far
 larger number of conventional resonances than the experiments which
 observe pentaquarks, and have much better mass resolution.
\begin{table}[htb]
\caption{\it Experiments with non-observation of the $\tht$ baryon.}
\centering
\vskip 0.1 in
\begin{tabular}{|c|l|c|c|}
\hline
Reference& Group& Reaction 	& Limit \\
\hline
\cite{bes} &
BES	& $e^+e^- \to J/\Psi \to \bar{\Theta}\Theta$	& $<1.1\times 10^{-5}$ B.R.  \\
\cite{babar} &
BaBar	& $e^+e^- \to \Upsilon (4S) \to pK^0 X$		& $<1.0\times 10^{-4}$ B.R.  \\
\cite{belle} &
Belle	& $e^+e^- \to B^0\bar{B}^0 \to p\bar{p}K^0 X$	& $<2.3\times 10^{-7}$ B.R.  \\
\cite{hera-b} &
HERA-B	& $p A \to K^0_S p X$		& $<0.02 \times \Lambda^*$  \\
\cite{sphinx} &
SPHINX	& $p C \to  \Theta^+ X$	& $<0.1 \times \Lambda^*$   \\
\cite{hypercp} &
HyperCP	& $\pi ,K,p Cu \to K^0_S p X$	& $<0.3\%\ K^0p$  \\
\cite{cdf} &
CDF	& $p \bar{p} \to K^0_S p X$	& $<0.03 \times \Lambda^*$  \\
\cite{focus} &
FOCUS	& $\gamma BeO \to K^0_S p X$	& $<0.02 \times \Sigma^*$   \\
\cite{belle-h} &
Belle	& $\pi ,K,pA \to K^0_s p X$	& $<0.02 \times \Lambda^*$   \\
\cite{phenix} &
PHENIX	& $Au+Au \to K^- \bar{n} X$	& (not given)   \\
\cite{lep} &
ALEPH	& $e^+e^- \to K^0_s p X$	&  $<0.07 \times \Lambda^*$   \\
\cite{compass} &
COMPASS	& $\mu ^+A \to K^0_s p X$	&  $-$   \\
\cite{l3} &
DELPHI	& $e^+e^- \to K^0_s p X$	&  $<0.5 \times \Lambda^*$   \\
\cite{e690} &
E690	& $pp \to K^0_s p X$	&   $<0.005 \times \Lambda^*$   \\
\cite{lass} &
LASS	& $K^+p \to K^+n\pi ^+$	&  $-$   \\
\cite{l3} &
L3	& $\gamma\gamma \to K^0_s p X$	&  $<0.1 \times \Lambda$   \\
\hline
\end{tabular}
\label{negative}
\end{table}
 The first significant negative result was published by the HERA-B
 collaboration~\cite{hera-b}.
 HERA-B does not see any evidence for the $\tht$ but
 observes a clear $\lamst$ and ${\overline{\Lambda}}$(1520) signals of about  2 thousand
 events.
 HERA-B obtains an upper limit on the ratio of
 production cross sections for the $\tht$
 and $\lamst$ of $ R_{\Lambda^*}<2.7\%$ at the 95\% CL for $M_{\tht}=1530$ MeV. 
 In the whole range of reported $\tht$ masses from $1522$ MeV to $1555$ MeV
 the limit varies up to 16\%.

 The ratio of  the $\tht$ and $\lamst$ production cross
 sections  $ R_{\Lambda^*}$ 
 is often used for the comparison of different experiments
 since $\lamst$ is narrow and easily reconstructed, it has a mass
 similar to the $\tht$ mass and one can draw similar diagrams for $\lamst$ and $\tht$
 production by exchanging an ${\overline{K}}$ meson into a $K$ meson. 
 The existence of similar
 diagrams unfortunately does not prove that production mechanisms for
 $\tht$ and $\lamst$ are similar. The ratio $ R_{\Lambda^*}$ 
 is of the order of unity in several
 experiments which observe the $\tht$  and less than a few
 percent in many experiments which do not see $\tht$ (see Table~\ref{negative}).

 In order to resolve this discrepancy many authors assume that the $\tht$
 production drops very fast with energy and is heavily suppressed in
 $e^+e^-$ annihilation. A model exists in which the $\tht$ production cross
 section is strongly suppressed at high energies in the fragmentation
 region~\cite{titov}. It is not clear how reliable this model is. In any case
 it is not applicable for the central production for example in the
 HERA-B experiment where some models predict the $\tht$ yield much higher
 than the experimental limits~\cite{thtprod}.
 
    However, the $\tht$ production mechanism is not known and therefore it
 is important to have a high statistics experiment at low energies
 where most evidence for pentaquarks comes from. This goal was achieved
 by the BELLE collaboration which analyzed interactions of low momentum
 particles produced in $e^+e^-$ interactions with the detector material.
 We will discuss this experiment after reviewing the situation with
 the anti-charmed and doubly strange pentaquarks.
 
\section{The anti-charmed pentaquark.}
       
The anti-charmed pentaquark was observed in the $p D^{*-}$ and
${\overline{p}}D^{*+}$ channels by the H1 collaboration both in DIS
and photo production~\cite{h1}.  After many experimental checks H1
concludes that the signal is real and self consistent. Still the
signal has very unusual properties. The $\thc$ measured width of
$(12\pm 3)$ MeV is consistent with the experimental resolution of
$(7\pm 2)$ MeV.  So its intrinsic width is very small although its
mass is $151$ MeV above the $p D^{*-}$ threshold and $292$ MeV above
$pD^{-}$ threshold. Its decay into $p D^{*-}$ is clearly visible
although naively one would expect much larger branching fraction for
the $p D^{-}$ channel where energy release is twice larger. Finally it
is produced with an enormous cross section. About 1.5\% of all charged
D* mesons are coming from decays of this new particle! These
properties are very surprising but we can not a priory exclude such a
possibility.
 
However, the ZEUS experiment which works at the same 
electron-proton collider HERA does not
see $\thc$ and gives an upper limit of 0.23\% at the 95\% CL on the fraction
of charged $D^*$ coming from $\thc$ decays~\cite{zeus-null}.
We denote this fraction  $R_{\thc/D^*}$. 
For DIS events with $Q^2>1$ GeV$^2$ the upper limit is 0.35\% at the 95\% CL.
 This is a clear
contradiction with the H1 result. We are not aware of any convincing
explanation of this discrepancy. One can try to explain the difference
using following  arguments. ZEUS detects more soft $D^*$ than H1. If
one assumes that pentaquarks are produced with high momenta only, than
$D^*$ mesons from their decays should be also energetic.
In this case soft $D^*$ that are more efficiently detected by ZEUS
should not be used in the comparison with H1.
However such an assumption does not resolve the discrepancy since ZEUS does not see
the signal also in the kinematic range very similar to the H1 one. 
 
The CDF collaboration also does not 
see any sign of $\thc$~\cite{cdf}. CDF has two orders of magnitude more reconstructed
$D^*$ mesons. They reconstruct $6247\pm 1711$  $D_2^{*0}\ra D^{*+}\pi^-$ and $3724\pm 899$
$D_1^0\ra D^{*+}\pi^-$ decays which have the event topology very similar to $\thc$. Majority of
charm particles at HERA and Tevatron are produced in the 
 fragmentation process. It is impossible to reconcile the results of
 the two experiments if $\thc$ is produced in the fragmentation process as
 well. No other mechanism was proposed so far.  There are also upper
 limits on $\thc$ production in $e^+e^-$ collisions by ALEPH~\cite{lep} and in photo
 production 
by FOCUS~\cite{focus}.           

We conclude that the evidence for $\thc$ 
is by far weaker than the evidence against it.

 \section{Doubly strange pentaquark.}

\begin{figure}[h!]
\centering
\begin{picture}(550,280)
\put(70,-10){\includegraphics[width=6cm]{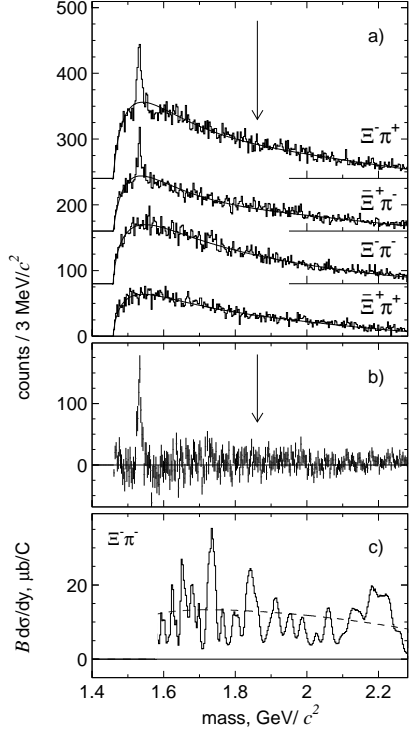}}
\end{picture}
\caption{\it The $\Xi\pi$ invariant mass spectra for $p+C$ collisions in the HERA-B experiment (a);
 sum of all four $\Xi\pi$ spectra with the background subtracted (b);
 upper limit at 95\%CL for mid-rapidity (c) .} 
\label{hera-b}
\end{figure}

The NA49 claim for the observation of the doubly strange pentaquark
was not supported by several experiments which tried to find
it. HERA-B has 8 times more  $\Xi^-$ hyperons and slightly better mass
resolution. There is no $\Xi(1862)$ signal in the $\Xi^-\pi^-$ or
 $\Xi^-\pi^+$
mass distributions (see Fig.~\ref{hera-b}) while there is a clear  $\Xi(1530)^0$ peak
with about 1000 events (including charge conjugate combinations). HERA-B
sets an upper limit of 4\%/$B(\Xi(1862)^{--}\ra\Xi^-\pi^-)$  at the 95\%CL on the ratio of
production cross section for  $\Xi(1862)^{--}$  and $\Xi(1530)^0$. We denote
this ratio $R_{\Xi (1862)/\Xi (1530)}$.
 $R_{\Xi (1862)/\Xi (1530)}$ is about 18\%/$B(\Xi(1862)^{--}\ra\Xi^-\pi^-)$
 in the NA49 experiment~\cite{hera-b,na49conf}.
  The center of mass energy in HERA-B is
about 2 times larger than in NA49. However the arguments about a very
fast drop of the pentaquark production cross section in the
fragmentation region~\cite{titov} do not apply to the central production
where the signal is observed by NA49~\cite{na49,note294} and where it is searched for at
HERA-B. The E690 experiment has even smaller limit on the $R_{\Xi (1862)/\Xi (1530}$  of
0.2\%/$B(\Xi(1862)^{--}\ra\Xi^-\pi^-)$ at the 95\% CL~\cite{e690}.
 E690 studies proton -
proton interactions at 800 GeV i.e. the same process as NA49 but at
the twice larger CM energy. The WA89 experiment has about 300 times
larger number of $\Xi^-$ hyperons but does not observe $\Xi (1860)$~\cite{wa89}.
 However this experiment uses a $\Sigma^-$ beam
and a straightforward comparison is not possible. The ALEPH, BaBar,
CDF, COMPASS, FOCUS and ZEUS experiments also  do not see $\Xi (1862)$ in a
variety of initial processes~\cite{lep,babar,cdf,compass,focus,zeus-null}.

 We conclude that the evidence for
$\Xi (1862)$ is by far weaker than the evidence against it.    
 
\section{The Belle experiment}

\begin{figure}[h!]
\centering
\begin{picture}(550,180)
\put(60,115){\rotatebox{90}{N/50 MeV}} 
\put(150,5){$p~(GeV)$} 
\put(55,-10){\includegraphics[width=8cm]{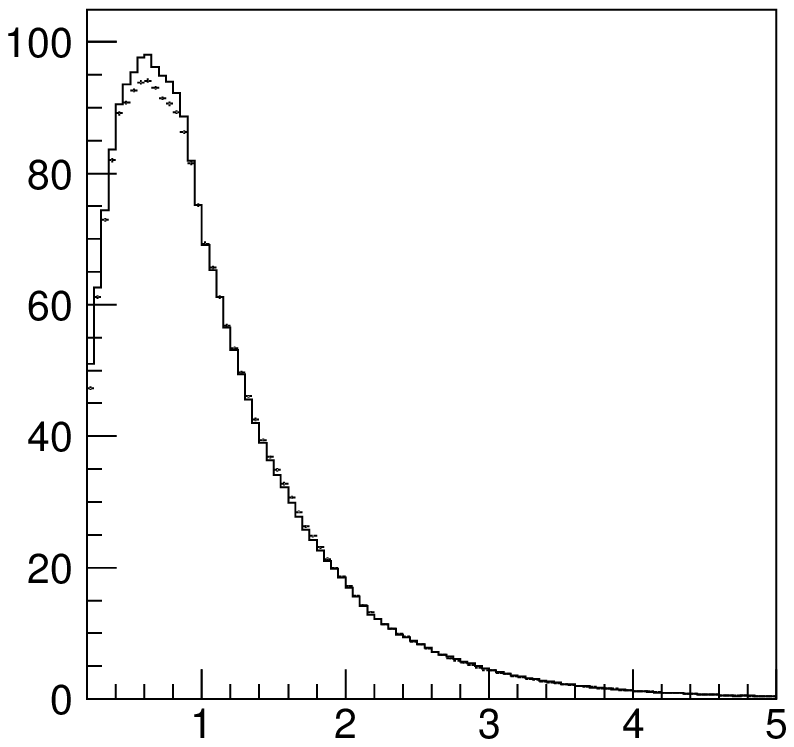}}
\end{picture}
\caption{\it Momentum spectra of $K^+$ (solid histogram) and $K^-$ (dashed histogram)
         in the Belle experiment.}   
\label{pk}
\end{figure}

As discussed above many high statistics experiments do not see the $\tht$ and
   set stringent limits on its production cross section in different
   processes. It was argued, however, that the $\tht$ production can be
   suppressed at high energies or in specific processes like $e^+e^-$
   annihilation. Therefore Belle decided to study interactions of low
   momentum particles produced in $e^+e^-$ interactions with the detector
   material. This allows to achieve production conditions similar to
   the experiments which observe the $\tht$ . For example the most probable
   kaon momentum is only 0.6 GeV (see Fig.~\ref{pk}). The Belle kaon
   momentum spectrum has a large overlap with the DIANA
   spectrum~\cite{diana}.

\begin{figure}[h!]
\centering
\begin{picture}(550,160)
\put(3,-10){\includegraphics[width=0.95\textwidth]{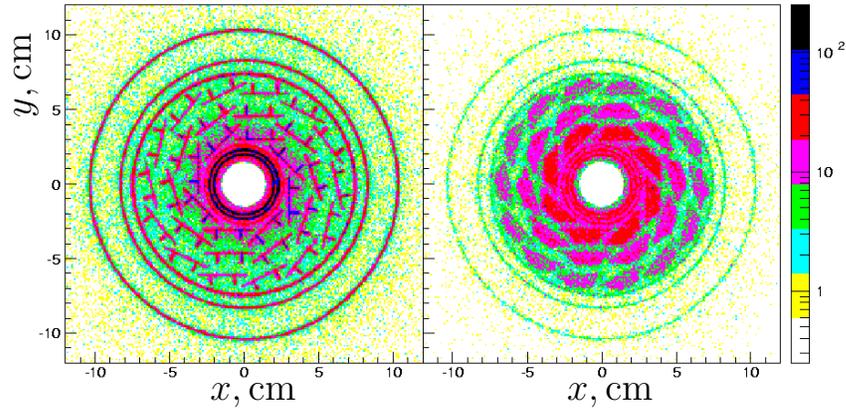}}
\put(11,110){\rotatebox{90}{\large $y,\cm$}}
\put(85,3){\large $x, \cm$} 
\put(220,3){\large $x, \cm$} 
\end{picture}
\caption{\it The $xy$ distribution of secondary $pK^-$ vertices 
for the barrel (left) and endcap (right) parts of the Belle detector.}  
\label{xy}
\end{figure}

   The analysis is performed by selecting $pK^-$ and $p\ks$
secondary vertices.  
The protons and kaons are required not to originate 
from the region around the run-averaged interaction point. 
The proton and kaon candidate are combined and the $pK$ vertex is
fitted. 
The $xy$ distribution of the secondary $pK^-$ vertices is shown in 
Fig.~\ref{xy} for the barrel part (left) and for the endcap part
(right) of the detector.
The double wall beam pipe, three layers of SVD, the SVD cover and the two
support cylinders of the CDC are clearly visible. The $xy$ distribution
for secondary $p\ks$ vertices is similar.  

\begin{figure}[h!]
\centering
\begin{picture}(550,180)
\put(60,115){\rotatebox{90}{N/5 MeV}} 
\put(125,5){$M(pK^-,p\ks),$ GeV} 
\put(60,-10){\includegraphics[width=8cm]{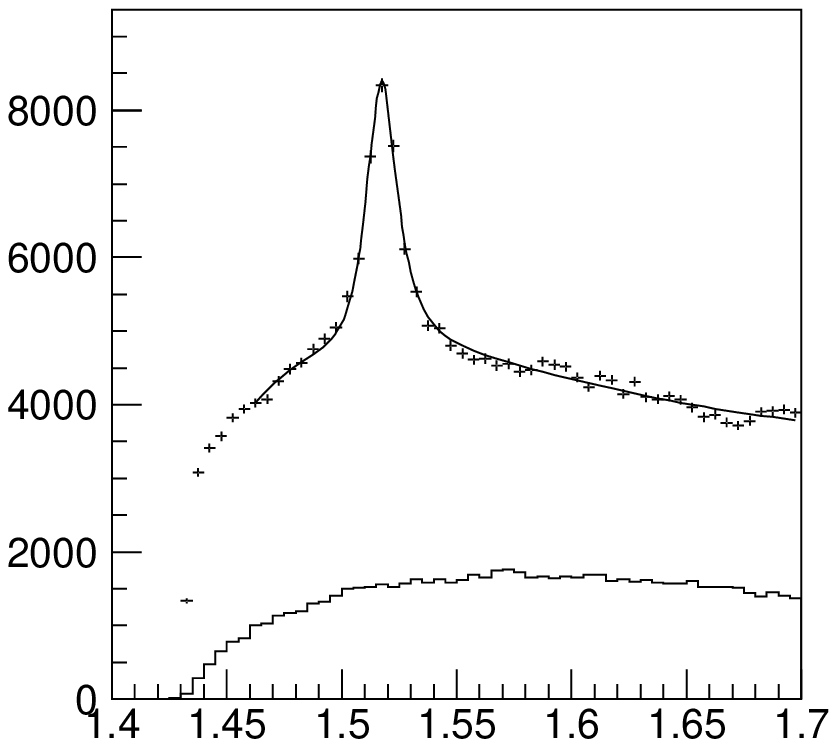}}
\end{picture}
\caption{\it Mass spectra of  $pK^-$ ( points with
  error bars) and $p\ks$ (histogram) secondary pairs in the Belle experiment.} 
\label{m_pk}
\end{figure}

The mass spectra for $pK^-$ and $p\ks$ secondary vertices
are shown in Fig.~\ref{m_pk}. No significant structures are observed
in the $M(p\ks)$ spectrum, while in the $M(pK^-)$ spectrum a
$\lamst$ signal is clearly visible. 

\begin{figure}[h!]
\centering
\begin{picture}(550,160)
\put(60,90){\rotatebox{90}{N/200 MeV}} 
\put(145,5){p~(GeV)} 
\put(60,-10){\includegraphics[width=7cm]{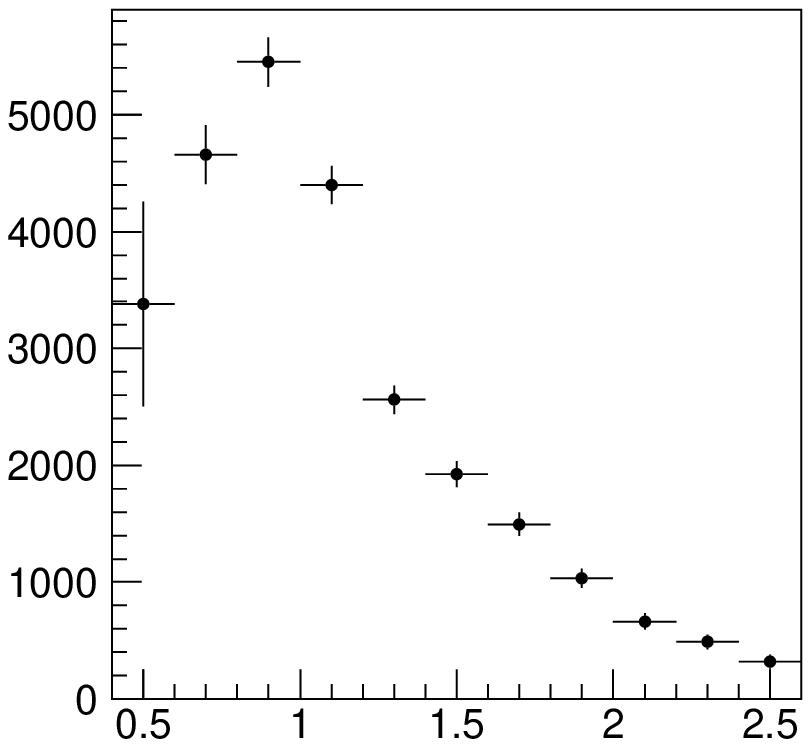}}
\end{picture}
\caption{\it $\lamst$ momentum spectrum in the Belle experiment.} 
\label{plam}
\end{figure}

The $\lamst$ yield is 15.5 thousand events. The $\lamst$ momentum spectrum
   is relatively energetic (see Fig.~\ref{plam}).
 $\Lambda (1520)$ 
produced in a formation channel should be contained mainly in the first bin 
of the histogram even in the presence of the Fermi motion. Therefore most of 
 $\Lambda (1520)$ are produced in the production channel.    

The upper limit for the narrow $\tht$ yield is 94 events at the 90\% CL for
  $M_{\tht}=1540$ MeV. This leads for the upper limit of 2\% at the 90\%CL on the
   ratio of $\tht$ and $\lamst$ production cross sections. For other reported
   $\tht$ masses the limit is even smaller.

 Projectiles are not
   reconstructed in the Belle approach. Therefore the $\tht$ and $\lamst$ can be
   produced by any particle originating from the $e^+e^-$ annihilation:
   $K^{\pm}$, $\pi^{\pm}$, $K^0_S$, $K^0_L$, $p$, $\Lambda$, etc.
    Belle shows that $\lamst$
   are seldom accompanied by $K^+$ mesons from the same vertex. This
   means that $\lamst$ are produced mainly by particles with negative
   strangeness. The fraction of energetic $\Lambda$ hyperons in $e^+e^-$
 annihilation is too small to dominate $\lamst$ production.

 The Belle 
limit is much smaller than the results reported by many
experiments which observe the $\tht$. 
For example it is two orders of magnitude smaller than the value
reported by the \mbox{HERMES} Collaboration~\cite{hermes}.
The $\tht$ and $\lamst$ are produced in  inclusive photoproduction at
   HERMES. Photons produce hadrons dominantly via (virtual) pions or Kaons. Therefore the
   production conditions are quite similar in the two experiments. We
   do not know any physical explanation for the huge difference
   between the Belle and HERMES results. 

   The expected number of reconstructed $\tht$ in the formation reaction 
   $K^+n\ra pK^0_S$
 can be
   estimated knowing the $\tht$ width, the number of $K^+$ mesons with
   appropriate momentum, amount of material and the reconstruction
   efficiency. The $\tht$ width was estimated using the DIANA data to be
   $0.9\pm 0.3$ MeV~\cite{cahn}.
Using this value of the $\tht$ width we estimate the number of expected
   $\tht$ events at Belle to be comparable with their upper limit. If so
   the Belle result disagrees with the DIANA observation. However we
   should wait for a quantitative statement from the Belle
   Collaboration. 

    A comparison of the Belle upper limit on $ R_{\Lambda^*}$  with the
   exclusive photoproduction experiments is not simple. However, it is
   very strange to have about two orders of magnitude difference in
   $ R_{\Lambda^*}$ since the Belle kaon (and pion) momentum spectrum is quite
   soft and comparable with the momentum spectrum of virtual kaons (or pions) in
   the low energy photoproduction experiments. 
     
\section{Conclusions.}

The NA49 claim for the observation of $\Xi(1862)$ pentaquarks is hard to reconcile with
the results of many experiments which have up to 300 times larger
statistics of usual $\Xi^-$ and $\Xi (1530)$ hyperons and a better mass
resolution. In particular E690 investigated the same production
process at about twice larger CM energy and obtained hundred times 
lower limit on the ratio of  $\Xi(1862)$  and $\Xi (1530)$ 
  production cross sections. 

   The H1 claim for the anti-charmed pentaquark contradicts the ZEUS study
   made at almost identical conditions.  CDF sets a very stringent
   limit on the $\thc$ yield although they observed 178 times more $D^*$
   than H1. CDF reconstructed also about 10 thousand $D_2^{*0}\ra D^{*+}\pi^-$
   and $D_1^{0}\ra D^{*+}\pi^-$  decays (including charge conjugate states).
   These decays are very
   similar in kinematics and efficiency to $\thc\ra pD^{*-}$ decays (the H1
   signal is observed mainly with energetic protons for which the
   particle identification does not play an important role). Three other
   experiments do not see any sign of the $\thc$ in different production
   processes~\cite{lep,belle,focus}.
 It is hard to reconcile the H1 claim with this
   overwhelming negative evidence. 

   The claims for observation of the $\tht$ in inclusive production at medium
   and high energies are not supported by many high statistics
   experiments which reconstruct by far larger number of ordinary
   hyperons with negative strangeness. Even if one assumes that the $\tht$
   production is strongly suppressed at high energies there is still a
   contradiction between several of these results with the Belle upper limit
   obtained with low momentum kaons.

    However, even if some claims for the  
   $\tht$ observation are wrong it does not mean that all observations are
   wrong. The DIANA and exclusive photoproduction experiments are not
   in contradiction with the high energy experiments if one assumes
   that the $\tht$ production drops very fast with the energy. There is a
   qualitative disagreement of these experiments with the Belle
   data. However here we should wait for the quantitative analysis of
   the Belle data. Results of high statistics exclusive
   photoproduction experiments are expected very soon.
 We hope that the situation with the
   pentaquark existence will be clarified already this year.

\section{Acknowledgements}
We are grateful to A.Dolgolenko, A.Kaidalov, R.Mizuk, and P.Pakhlov for the fruitful discussions.

\newcommand{\etal}{ {\it et al.}, }

\end{document}